\documentclass[12pt]{amsart}
\usepackage[margin=1in]{geometry}

\date{\today}
\usepackage{mathtools}
\usepackage{graphicx}
\usepackage{multicol}
\usepackage[inline]{enumitem}
\usepackage{graphicx}
\usepackage{tikz-cd}
\usepackage{amssymb}
\usepackage{relsize}
\usepackage{mathdots}
\usepackage{arcs}
\usepackage{mathrsfs}
\usepackage{bbm}
\usepackage{pgf,pgfarrows,pgfnodes,pgfautomata,pgfheaps,pgfshade}
\usepackage{tikz}
\usetikzlibrary{patterns,decorations.pathreplacing,calc,intersections}

\setenumerate[1]{label={\textbf{Problem \arabic*}},leftmargin=0cm,labelwidth=\itemindent,labelsep=.3cm,align=left,itemsep=.5cm}\setenumerate[2]{label={\textbf{(\alph*)}\ \ },leftmargin=0cm,labelwidth=\itemindent,align=left,labelsep=.1cm,itemsep=0pt}
\theoremstyle{plain}

\theoremstyle{definition}
\theoremstyle{remark}

\begin{document}

\pagebreak
\thispagestyle{plain}
\section*{\textbf{\larger[2]Combinatorial Nullstellensatz Techniques}}
\section*{daniel l. freed\\university of chicago}

\section*{}
\section*{\textbf{Abstract}}
We present different techniques for applying Combinatorial Nullstellensatz (CNSS) to polynomials over finite fields. For examples, we generalize theorems from Noga Alon's paper on the subject [1], and present a few of our own.
\section*{\textbf{Introduction}}
\noindent\textbf{Theorem 1.}       \textit{(Combinatorial Nullstellensatz \textbf{[1]})  } 
Let $F$ be a field, and let $f=f(x_1,\dots,x_n)$ be a polynomial in $F[x_1, \dots, x_n]$. Suppose the degree of $f$ is $\sum\limits^n_{i=1}t_i$ where each $t_i$ is a nonnegative integer, and also suppose that the coefficient of $\prod\limits^n_{i=1}x_i^{t_i}$ in $f$ is nonzero. Then, if $S_1,\dots, S_n$ are subsets of $F$ with $\lvert S_i \rvert > t_i$, then there are $s_1 \in S_1, s_2 \in S_2, \dots s_n \in S_n$ so that 
$f(s_1,\dots, s_n) \neq 0$
\\\\
We will occasionally use a generalization of this theorem, from a paper by Michael Lason [2], with a weaker assumption on the degree of the nonvashing monomial. Let $F$ be a field. For $f\in F[x_1,...,x_n]$, define: $$Supp(f) = {\{(a_1,...,a_n) \in N^n: \text{the coefficient of} \prod\limits_{i=1}^n x_i^{a_i} \text{is non-zero}\}}$$
Where $Supp(f)$ has the natural partial ordering: $(a_1,...,a_n)\geq(b_1,...,b_n)$ if and only if $a_i \geq b_i$ for all $i$.\\
\\\textbf{Theorem 2.}\textit{ (Generalized Combinatorial Nullstellensatz \textbf{[2]})}\textit{ Let $F$ be a field and f a polynomial in $F[x_1,...,x_n]$. Suppose $(a_1,...,a_n)$ is maximal in $Supp(f)$. Then for any subsets $A_1,...,A_n$ of $F$ such that $|A_i| \geq a_i + 1$, there are $a_1 \in A_1,..., a_n \in A_n$ such that $f(a_1,...,a_n) \neq 0$ }

\section*{\textbf{1. Excluding a Value}}
Suppose we have a set of elements that satisfy some property. If we can represent this set as the roots of a polynomial, applying Combinatorial Nullstellensatz can tell us if and when there exist elements that don't satisfy the property. However, there might be a trivial element that we want to exclude from our search. \\
\\
To do so, say we intend to imply CNSS on a polynomial $f$. To exclude some value, $a$, create an exclusion polynomial $g$ such that $g(a) = -f(a)$, and for all other $x, g(x) = 0$. Then applying CNSS
to $(f+g)$, the existence of a non-zero value implies that it can’t be $a$, since $(f+g)(a) = 0$, but also
that $f$ must be non-zero, since $g$ is 0 everywhere else. \newline
\\
In certain cases, excluding a value can increase the degree of our polynomial; so although we have restricted our view on the subsets
$S_i$, they may have to be larger than they were before. But in many situations, we may not initially have a decent leading term to apply CNSS on, in which case this method can help if you have values to exclude.\\
\\
To extend this, we can also exclude multiple values, either with a single polynomial\footnote{see Section 1.2. "A Different Approach"} or by reiterating the above process\footnote{Be sure to pay careful attention to how the leading terms of the exclusion polynomials interact.}. 
\\\\
For an example, here is Theorem 3.1 from Noga Alon's paper [1].
\\ \newline
\textbf{Theorem 3.} \textit{(Chevalley)} \textit{ Suppose we have $m$ polynomials $P_1,\dots, P_m \in F_p[x_1,\dots, x_n]$. If they share a common zero, and
the sum of their degrees is less than n, then they share another one.} \\\\
\textit{Proof.}
We would like to create a
polynomial $f(x_1,\dots, x_n) \in F_p[x_1,\dots,x_n]$ which goes to $0$ on all inputs except when that input is a common root of $P_1$ through $P_m$. Since roots by definition send the polynomials to zero, we
want to take the negation of the output of a polynomial mod $p$.
We do so using Fermat's Little Theorem:
\begin{align*}
1-(P_i(x_1,\dots,x_n))^{p-1}
\end{align*}
Now this is non-zero only on roots of $P_i$. So to make our polynomial non-zero only on a common
root, we take the "and" of these statements:
\begin{align*}
f(x_1,\dots, x_n) = \prod \limits _{i=1}^m[1-(P_i(x_1, \dots, x_n))^{p-1}]
\end{align*}
$f$ is already non-zero by assumption, so we would like to exclude the value of the common root $c
= (c_1,\dots,c_n)$. \\ 
\\
To do so, take the polynomial $\prod\limits_{j=1}^n \prod \limits_{q \in F_p, q\neq c_j}
(x_j - q).$
This sends everything to $0$, except the root. Currently, on the value c, this goes to ${(-1)}^n$. Since $f(c) = 1$ we want this exclusion polynomial to go to $-1$ on $c$, so we let \\
$$g = -(-1)^n\prod\limits_{j=1}^n \prod \limits_{q \in F_p, q\neq c_j}
(x_j - q).$$ By assumption $\sum\limits_{i=1}^m deg(P_i) < n$, so $g$ has larger degree than $f$. Then applying CNSS to $f+g$ using the leading term $\prod \limits_{j=1}^n {x_j}^{p-1}$ completes the proof.\footnote{Alon shows that for multivariate polynomials, a good way to exclude values is to do so component-wise. (As long as the value you are excluding attains a specific known value in the original polynomial, this exclusion polynomial serves as a general tool for excluding a specific input over a finite field.)} \qed \\

\section*{\textbf{1.1. A Generalization of the Chevalley-Warning Theorem}}
Now that we know there is another root, $d=(d_1,\dots,d_n)$, we can exclude it from our search as well.  After adding on the exclusion polynomial
 $h = -(-1)^n\prod \limits_{k=1}^n \prod \limits_{q \in F_p,q \neq d_k}
(x_k - q)$,\\
\vspace{-.32cm}

\noindent our leading term, $\prod \limits_{i=1}^n{x_i}^{p-1}$, will
have leading coefficient -2 or 2 for $n$ odd or even. For $p>2$ this is non-zero, so applying Nullstellensatz to $f+g+h$, we have that our polynomial has even another
root. \\
\\
Repeating this process shows that the amount of shared roots of the polynomials is a multiple of $p$.
It turns out that this is the same result as Warning's extension of Chevalley's theorem from 1935.\footnote{Everything we have done easily generalizes to an arbitrary finite field of order $p^k$. The product of the units is still $-1$, and the field is still characteristic p.} \\
\\
Now, let's see what happens in the case where $n= \sum \limits_{i=1}^m deg(P_i)$
\\ \newline
\textbf{Theorem 4.}
\textit{Take $P_1,\dots,P_m \in F[x_1,\dots,x_n]$, where $F$ is a finite field of order $p^k$. 
Suppose that $\sum \limits_{i=1}^mdeg(P_i) \leq n$.
Further suppose that each of the $P_i$ is $\neq 0$.\footnote{Clearly, if one is 0 the amount of shared roots is the amount the rest share.} \\ \newline
Then if the $P_i$ are non constant, and $\prod \limits_{i=1}^m P_i$ contains $q\prod \limits_{i=1}^n x_i$ as a leading term (for $q \neq 0$), then the amount of shared roots of the $P_i$ is 1 mod $p$ if the parities of $n$ and $m$ are the same, and -1 mod $p$ if
the parities of $n$ and $m$ are different. \\ \newline
Otherwise (if one of the $P_i$ is constant or the product doesn't contain the necessary leading term)
the amount of shared roots is 0 mod $p$.} \\ \newline
\textit{{Proof.}} 
It is easily checked that $\prod \limits_{i=1}^m(1 - (P_i)^{{p^k}-1})$ will have
leading term $(-1)^m \prod \limits_{j=1}^n {x_j}^{{p^k}-1}$ if and only if each $P_i$ is non
constant and $\prod \limits_{i=1}^m P_i$ has leading term $q \prod_{i=1}^n x_i$ for
$q \neq 0$. \\
Then notice, that if $\prod \limits_{i=1}^m(1 - (P_i)^{{p^k}-1})$ has
leading term $(-1)^m \prod \limits_{j=1}^n {x_j}^{{p^k}-1}$, the amount of shared roots
is 1 mod $p$ if the parities of $n$ and $m$ are the same and -1 mod $p$ if their parities are different. This is just using the same repetition argument, noticing that if the parities are the same, then the leading term from the product will be the additive inverse of the leading term of an exclusion polynomial, and otherwise the leading terms will be the same. \\\\
If $\prod \limits_{i=1}^m(1 - (P_i)^{{p^k}-1})$ does not end up having that leading term, then
there will be no interaction with our exclusion polynomials (the exclusion polynomials will
contain a variable that it doesn't have), so again by the repetition argument, the amount of shared roots is 0 mod $p$. \qed\\
\\Let's see the case where $\sum\limits_{i=1}^n deg(P_i)>n$; 
\\\textbf{Theorem 5.} \textit{Let $P_1,...,P_m \in F[x_1,...,x_n]$, where F is a finite field of order $p$. Suppose $\sum\limits_{i=1}^m deg(P_i) > n.$ Let $f= \prod\limits_{i=1}^m (1 - P_i(x_1,...,x_n)^{p-1})$ and reduce the degrees of its monomials mod(p-1) to obtain a reduced polynomial $g$ with equivalent evaluation map. Suppose the term $\prod\limits_{j=1}^n {x_j}^{p-1}$ of $g$ has leading coefficient $d$ (where $d$ may possibly be 0). Then the amount of shared roots is $(-1)^n d$ (mod p).}
\newline\newline
\textit{{Proof. }} If n is even, then the leading term of our typical exclusion polynomial is $-\prod\limits_{j=1}^n {x_j}^{p-1}$. Thus repetitive application of Generalized CNSS yields that the amount of roots is \newline {$d$(mod $p$)}, as we need $d$ exclusion polynomials in order to initially cancel out the term. Similarly, if n is odd, then the leading coefficient of the term will be positive, and thus we will need $-d$ exclusion polynomials to initially cancel out the term. 

\section*{\textbf{2. General Corollaries of Exclusion Polynomials}}
By continuing to apply different exclusion polynomials to an arbitrary polynomial $f$, we can determine general corollaries that can be applied to more specific cases.\\\\
\textbf{Theorem 6. }\textit{Let $f \in F_p[x_1,...,x_n]$ Define $S$ to be the subset of ${\{0,1\}}^n$ such that $f\neq 0$ (resp $f=0$). Suppose no term of $f^{p-1}$ contains all $x_i$, for $i = 1,...,n$. Then the amount of elements in $S$ with an even number of ones equals the amount with an odd number of ones $modulo$ $p$.}\\
\\ \textit{Proof.} Let $g = f^{p-1}$. Consider the exclusion polynomial: $$h = (-1)^{k+1}\prod\limits_{i=1}^n (Q_i-x_i)$$ where $Q_i$ is the complement of an input from ${\{0,1\}}^n$ such that g is non-zero, and $k$ is the amount of ones in the input. Then as long as no term of $f^{p-1}$ contains all $x_i$, we can apply Generalized CNSS to $g$ iteratively using the exclusion polynomial $h$. Notice that for a given excluded input, the sign of our exclusion polynomial changes depending on the parity of the amount of ones in the input. Thus, the amount of elements in $S$ with an even number of ones equals the amount with an odd number of ones $modulo$ $p$. That is, suppose the amounts were not equal. Then after excluding all elements of S, we would have a leftover exclusion polynomial $h$ that was not cancelled out. Thus, there would be another element of S, which is a contradiction.  Applying the same exclusion polynomial to $g = 1 - f^{p-1}$ completes the same proof in the case where $S = {\{ x\in {\{0,1\}}: f(x)=0\}}$. $\qed$ \\
\textbf{Corollary:} \textit{Let $ f_1,...f_k \in F_p[x_1,...,x_n]$, and let $g = \prod\limits_{i=1}^k (1-f_i^{p-1})$ Suppose g does not contain a term of the form $\prod\limits_{i=1}^n x_i^{k_i}$, where $k_i$ is non-zero for $i = 1,..,n.$ Then we have that the amount of shared roots of $f_1,...,f_k$ in ${\{0,1\}}^n$ with an even number of ones, equals the amount with an odd number of ones \textit{modulo p}.}\\
\\\textbf{Corollary:} \textit{Let A be a set. Let $f_1,...,f_k \in F_p[x_1,...,x_{|A|}]$ We say that $S\subseteq A$ satisfies $f_1,...,f_k$ if the indicator vector for $S$ is a shared root of $f_1,...,f_k$. Then, provided that $(p-1)\sum\limits_{i=1}^k deg(f_i) < |A|$, then the amount of even subsets that satisfy $f_1,...f_k$ equals the amount of odd subsets that satisfy them modulo p.}\\ 
\newline
\textbf{Theorem 7.}
\textit{Let $f \in F_p[x_1,...,x_n]$ Define $S_o$ to be the subset of ${\{0,1\}}^n$ with an odd number of ones, such that $f\neq 0$. Define $S_e$ to be the subset of ${\{0,1\}}^n$ with an even number of ones, such that $f\neq 0$. Applying Lagrange's theorem, reduce the degrees of terms of $f^{p-1}$, mod($p-1$), to obtain a reduced polynomial $g$ with equivalent evaluation map. Then, if $g$ contains the term $d\prod\limits_{i=1}^n x_i$}, then we have that $|S_e| - |S_o| = (-1)^nd$ (mod $p$).
\newline
\\\textit{Proof.}
Suppose n is even, then applying the same exclusion polynomial from Theorem 6, in order to cancel out the term, we would need to add on $d$(mod p) more exclusion polynomials associated with an input with an even number of ones than ones associated with odd with an input with an odd number of ones. The argument is symmetric in the case that n is odd.\\ 
\\\textbf{Theorem 8:} \textit{Let $f \in F_p[x_1,...,x_n]$, and $f\neq 0$. If (reduced) f doesn't have a term that dominates $\prod\limits_{i=1}^n x_i^{p-1}$, then f has multiple non-zero values.}\\
\\\textit{Proof.}
Let $f \in F_p[x_1,...x_n]$, with $f(c_1,...,c_n) = d \neq 0$ and consider the exclusion polynomial$$g = -d(-1)^n\prod\limits_{j=1}^n \prod \limits_{q \in F_p, q\neq c_j}(x_j - q)$$ Then as long as f doesn't have a term that dominates $\prod\limits_{i=1}^n x_i^{p-1}$, then f has  another non-zero value.\\
\\Then by reduction, we have that the only polynomials that can have a single non-zero value must have that term. Now suppose our polynomial has the term. Then, applying CNSS, for some $c\in F_p^n$ we have $f(c)=d\neq 0$. Then, excluding that value with the same polynomial $g$,  we have that $\prod\limits_{i=1}^n x_i^{p-1}$ has leading coefficient $d(-1)^{n+1}$. Thus the coefficient of the $\prod\limits_{i=1}^n x_i^{p-1}$ in $f$ must be $d(-1)^n$ if our polynomial is to have a single non-zero value, since otherwise applying CNSS again would yield another one.\\

\section*{\textbf{2. Expressions in $F_p$}}

Let $J$ denote some expression that counts an amount dependent on input variables. Then
$f = 1 - (J-k)^{p-1}$
 being non zero implies $J$ is congruent to $k$ mod $p$ on some input value. For an example of how this can be used, here is a theorem based off of ideas from Theorem 6.1 in [1].\newline
\\
\textbf{{Theorem 9}}: \textit{Given a graph $G = (V,E)$, if $|V|(p-1) < |E|$, then for every k, the amount of even subsets K of E such that the degree with respect to K of every vertex $v \in V$ is k(mod p) is equal to the amount of odd subsets that satisfy the same property modulo p.}\newline 
\\
\textit{Proof.}
We construct a polynomial from $F_p[x_1,...,x_{|E|}]$, and show the existence of subsets of edges by using CNSS to choose an input from ${\{0,1\}}^{|E|}$. Letting our expression be $J = \sum \limits_{e \in E}(a_{v,e} x_e)$, where $a_{v,e}$ is the incidence matrix for our graph, gives us:

\begin{align*}
f = \prod \limits_{v \in V}[1-(\sum \limits_{e \in E}(a_{v,e}x_e)-k)^{p-1}]
\end{align*}
\\
We take the product over all $v \in V$ so that our expression holds for each of them. For an input representing a subset of edges, $J$ counts the amount of edges in our subset that are adjacent to the specified vertex $v$. Our polynomial will be non-zero if and only if this amount is equal to $k$(mod $p$), for every $v$ in $U$, implying we have a subset of edges such that the degree of every vertex in $U$ under this subset is $k$(mod $p$). Then, applying Theorem 6 completes the proof, since the leading term of $f$ is smaller than the amount of variables. \qed\\\\ 

In this case, our expression counted the degree of a given vertex with respect to a subset of edges. It's then natural to flip this around and count the degree of a given vertex with respect to a subset of vertices, doing so by counting the amount of edges from the subset to the specified vertex. \\\\
For $G=(V,E)$, let $U\subseteq V$ and consider the polynomial:

$$\prod \limits_{u \in U} 1-\left((x_u \sum \limits_{v \in V} e(u,v)x_v) - k\right)^{p-1}$$
where $e(u,v)$ is 1 if $(u,v)\in E$ and is 0 otherwise. 

If $|V| > 2(p-1)|U|$, and ($k \neq 0$), then we can use this to study the amount of subsets of vertices containing U such that the amount of edges between each member of U and elements of the subset is k(mod p). \\

Provided that $|V| > 2$ or $p > 2$ this also holds for $|V| = 2(p-1)|U|$ (check to make sure this makes sense)\\
\\To exclude a subset of vertices, we would use the polynomial $g= \prod \limits_{v \in V} (x_v-Q_v)$, where $Q_v$ is the indicator vector for the complement of our subset. Notice, that as in Theorem 3, the sign of this exclusion polynomial depends on the parity of the size of the excluded subset. This is inconvenient for ease of excluding multiple values, but gives us the additional information that the amount of even subsets satisfying the condition must equal the amount of odd subsets satisfying the condition $modulo$ $p$. (which could be a theorem I guess). If it didn't, then there would exist another subset that satisfies the condition contradicting the fact that all were exhausted. In fact, if we take $U=\varnothing$, $k=0$, and consider a graph with no edges for a given choice of V, we have an alternate proof that the amount of even subsets of an n-set equals the amount of odd subsets. This is just a more contrived way of excluding values off of the identity polynomial. I think interesting theorems could be found by considering different examples of graphs and excluding values. \\
\\Remark: We can of course alter this polynomial in a variety of ways to create other similar ones with interesting meanings. For example, $$\prod \limits_{u \in U} 1-\left(\sum \limits_{v \in V} e(u,v)(1 - x_v) - k_{v,e}\right)^{p-1}$$
or
$$\prod \limits_{u \in U} 1-\left(\sum \limits_{v \in V} e(u,v)(1 - x_u)x_v - k_{v,e}\right)^{p-1}$$\\
\\\textbf{Proposition 6.2 from [1]:} \textit{Let p be a prime and let $G=(V,E)$ be a graph such that $|V|> d(p-1)$.\footnote{Can have equality provided $|V| \neq d$ or V is not complete} There is a non-empty $U \subseteq V$ such that the number of cliques of d vertices of G that intersect U is $0$ $modulo$ $p$.}\\\\
\textit{Minor Generalization and Comments:}
To show this, for each $I \subseteq V$, let $K(I)$ denote the number of copies of $K_d$ in $G$ that contain $I$. Let $f \in F_p[x_1,...,x_{|V|}]$ be given by $$f = 1 -\left(\sum \limits _{I \subset V, I \neq \emptyset} (-1)^{\lvert I\rvert+1}K(I)\prod \limits _{i \in I} x_i - k\right)^{p-1}$$ Then, by Inclusion-Exclusion, f is non-zero on an indicator vector for a subset is non-zero if and only if the amount of $K_d$ that intersect $U$ is $k$ $modulo$ $p$. Supposing there exists a subset U that satisfies this property, we can exclude it with $$(-1)^{|U|+1}\prod\limits_{v \in V}(Q_i - x_i)$$
Then the proposition follows from letting $k=0$ and excluding the empty set.\\\\
Notice that by the exclusion polynomial's dependence on the parity of the excluded subset, we have that the amount of even subsets satisfying the property equals the amount of odd subsets satisfying it, $modulo$ $p$. Then for example, in the case where $k=0$, if we have a non-empty subset with even cardinality satisfying the property, then after excluding both it and the empty set, we have that there exist at least two more subsets satisfying it.\\

\section*{\textbf{1.2. A Different Approach}}
To give another example of excluding a value, look at the polynomial $f(x,y) = xy - (x+y)$ over an
arbitrary field, $F$. Then applying CNSS, we have that given any two subsets of F of size 2, then there exists an element from each of them such that their sum doesn't equal their product. \\ \\
This is already a bit more interesting than it seems. For example, taking these subsets to be the same implies: for any $a,b \in F$, one of the combinations $(a,a), (a,b)$, or $(b,b)$ has
the property that the sum of the two elements doesn't equal it's square. This gives you
identities like, if $a^2 = 2a$ and $b^2 = 2b$ then $a+b \neq ab$, and whatever others you can think of.\\ \\
But we might as well see what we can find by excluding some values. Restricting our attention back to finite fields, to exclude all inputs of the form $(a,0)$ and $(0,b)$, add on the polynomial \\
$$g(x,y) = x(x^{p-1} - y^{p-1}) + y(y^{p-1} - x^{p-1})$$ \\
Then, $g(0,0) = 0$.
For $x,y \neq 0$, $g(x,y) = 0$ since both the left and right sides get sent to 0.
For $x \neq 0$, $g(x,0) = x$ since the LHS is $x$ and the RHS is 0.
Similarly, for $y \neq 0$, $g(0,y) = y$ \\ \\
Since $f(x,0) = -x$ and $f(0,y) = -y$, $g$ excludes these values.\footnote{This may look a little contrived, but the intuition for the construction is just that $x^{p-1} - y^{p-1}$ tells you when one input is 0 and the other isn't (and which input from the sign).} \\\\
To additionally exclude pairs of additive inverses, add on the polynomial $h(x,y) = (1-(x+y)^{p-1})x^2$. (Can maybe do -xy instead here). It excludes additive inverses since WLOG for $a \neq -b$, $h(a,b)$ gets sent to 0. And for $a = -b$, $h(a,b)$ goes to $a^2$, which cancels out $(f+g)(a,b)$. \\ \\
Looking at the polynomial $(f+g+h)(x,y)$,
choose some leading term of the form $x^{p-k+1}y^k$ with non-zero leading coefficient $-\binom{p-1}{k}$ and then applying CNSS: take any $A,B \subseteq F_p$ with $\lvert A \rvert = p-k+2$ and $\lvert B \rvert = k+1$, for $k$ an element of $\mathbb{Z}^+$. Then there exists $a \in A$ and $b \in B$ such that $a + b \neq ab$. And additionally, $a$ and $b$ are not additive inverses and neither one of them is 0. \\
\\This highlights a trade-off of excluding a value. Although we added restrictions to our search, we increased the size of the set we were searching in.

\newpage
\section*{\textbf{References}}
\noindent [1] Noga Alon, Combinatorial Nullstellenstatz; Tel Aviv University, January 1999 \newline\newline
[2] Michael Lason, A Generalization of Combinatorial Nullstellensatz, 2013 \\

\end{document}